\documentclass[12pt]{article}

\usepackage[dvips]{graphicx}
\usepackage{amsfonts,amsmath,amssymb,epsf,color,graphics}

\usepackage{epsfig}
\usepackage[dvipdfmx,hypertex]{hyperref}

\newcommand{\ba}{\begin{eqnarray}}
\newcommand{\ea}{\end{eqnarray}}

\def\m{{\mu}}

\def\Om{{\Omega}}

\def\l{{\lambda}}
\def\G{{\Gamma}}

\def\s{\sqrt}

\def\al{\alpha'}

\def\f {\frac}
\def\ti{\tilde}

\begin{document}

\begin{titlepage}
  \thispagestyle{empty}
    
  \begin{flushright}
    
    KUNS-2075
    
    RIKEN-TH 101
  \end{flushright}
  
  \bigskip
  
  \begin{center}
    \noindent{\LARGE{Cascade of Gregory-Laflamme Transitions \\ 
    \vspace{-0.1cm}
	and \\
	\vspace{0.3cm}
	$U(1)$ Breakdown in Super Yang-Mills}}\\
    
    \vspace{2cm} 
    
     \noindent{
       Masanori Hanada$^a$\footnote{e-mail:hana@riken.jp} and 
       Tatsuma Nishioka$^b$\footnote{e-mail:nishioka@gauge.scphys.kyoto-u.ac.jp}}\\
    
    \vspace{1cm}	
    
	 $^a$ {\it Theoretical Physics Laboratory,
RIKEN Nishina Center,

 Wako, Saitama 351-0198, Japan
} 	\\
    
     $^b$ {\it  Department of Physics, Kyoto University, Kyoto 606-8502, Japan }
    
    \vskip 3cm
  \end{center}

  \begin{abstract}
      In this paper we consider black $p$-branes on square torus.
   We find an indication of a cascade of Gregory-Laflamme transitions
   between black $p$-brane and $(p-1)$-brane.
   Through AdS/CFT correspondence, these transitions 
   are related to the breakdown of the $U(1)$ symmetry in super Yang-Mills on 
   square torus.
   We argue a relationship between the cascade and recent Monte-Carlo data.
     
  \end{abstract}
  
\end{titlepage}

\newpage

\section{Introduction}
\setcounter{equation}{0}
In general relativity a black string becomes unstable and changes to 
a black hole when the circle it winds becomes large compared with its horizon.  
This is known as the Gregory-Laflamme transition \cite{GrLa} (see for review \cite{Ko,HNO}). 
Similar instability exists also for neutral black $p$-brane wrapped on torus. 
In the case of square torus, in which all compactification radii are the same, 
usually only a black hole and a non-uniform brane are 
considered as candidates of the final state, probably because they 
respect the symmetry of the background spacetime. 
However, there is no reason for adhering them - 
isotropy can break spontaneously, and we should also consider 
a decay from $p$-brane to $(p-1)$-brane.   

A motivation for considering such a decay 
comes from AdS/CFT correspondence \cite{Ma,GKP,Wi}. 
Through AdS/CFT, 
the Gregory-Laflamme transition is related \cite{Su,BKR,MaSa,MaSa2,AMMW,HO} to the condensation of 
the spatial Wilson loop winding on compact direction \footnote{In the following 
we call it simply ``spatial Wilson loop''. } 
in dual super Yang-Mills theory. 
Recently, Narayanan, Neuberger and collaborators have studied 
phase structure of pure bosonic Yang-Mills theory on torus intensively \cite{NN}. 
According to their result, the spatial Wilson loops  
condensate not simultaneously 
but one-by-one even if compactification radii are the same. 
We can expect similar pattern of successive phase transitions 
in bosonic Yang-Mills on $T^n$ with $10-n$ adjoint scalars, which is 
the high-temperature limit of the super Yang-Mills on $T^{n+1}$. 
Then, it is plausible that dual supergravity theory also has a cascade
 of Gregory-Laflamme transitions. In this paper, we compare 
the free energies for several solutions in (super)gravity using a certain 
approximation. The result suggests that such a phase structure exists. 
In IIA supergravity the free energy of D$p$-brane\footnote{Precisely speaking, 
smeared D$0$-branes obtained via T-dual.} 
becomes smaller than that of D$(p-1)$-brane at some critical temperature 
$t_{C(p)}$, which satisfies $t_{C(p)} > t_{C(p-1)}$. 
In the case of the transition from D$1$- to D$0$-brane, the Gregory-Laflamme 
temperature appears slightly below  $t_{C(1)}$. 
This implies that the D$1$-brane collapses to the D$0$-brane. 
We expect a similar pattern for any $p$. 
In Einstein gravity,  we confirm this pattern explicitly
for the neutral black branes 
(see Table \ref{tab:1}, Table \ref{tab:1.5} and Figure~\ref{fig:1}). 

The organization of this paper is as follows. 
In \S\ref{sec:Scwarzschild} we consider the neutral black branes.
Approximating compact directions transverse to brane with noncompact ones,
 we find an indication that $p$-brane does not decay to $0$-brane directly, but
its dimension decreases one-by-one, that is, $p$-brane decays to 
$(p-1)$-brane, then to $(p-2)$-brane, and so on.
In \S\ref{sec:SYM}, we summarize basic properties of weakly-coupled Yang-Mills theory,
and then we study the phase structure of dual supergravity 
in \S\ref{sec:gravity}. We find similar phase structures in both sides.
This result might be an  evidence of AdS/CFT, 
although the coupling region we considered are different in both sides.
\S 5 is devoted to discussions.
\section{Neutral black branes on ${\mathbb R}^{D-n}\times T^n$}
\label{sec:Scwarzschild}
\setcounter{equation}{0}
In this section, we calculate the free energy of neutral black $p$-branes 
winding around square torus $T^n$ and   
 evaluate critical temperatures $t_{C(p)}$, above which $(p-1)$-brane has 
smaller free energy than $p$-brane.
We also calculate the Gregory-Laflamme (GL) temperature $t_{GL(p)}$, 
above which $p$-brane becomes unstable.
In these calculations, we approximate compact directions transverse to brane 
by noncompact ones because of the lack of analytic solution for black branes
in compact space.
We find that 
\begin{equation}\label{eq:ordering}
\dots < t_{C(p)} < t_{GL(p)} < t_{C(p-1)} <  t_{GL(p-1)} < \dots
\end{equation}
for $D \le 12$ and $D-p \ge 4$ which indicates the existence of a cascade of 
first order transitions\footnote{In the case of  $D \ge 13$, 
(\ref{eq:ordering}) does not hold. For example, at $D=13$, we obtain 
$t_{GL(p)} < t_{C(p)}$ for $p \le 3$. Although this is consistent with 
\cite{KS}, however, our approximation becomes subtle at such large dimensions
as pointed out in \cite{KoSo2}. More careful treatment is needed in order to 
 show whether or not our approximation could be applicable at larger 
dimensions.
}.
\subsection{Preliminary}
Let us consider the $D$-dimensional Einstein-Hilbert action 
in the asymptotically flat space, 
\begin{equation}\label{eq:DdimEH}
I_{EH} = \f{1}{16\pi G_N^{(D)}}\int d^Dx \s{-g}R.
\end{equation}
We take the background spacetime to be ${\mathbb R}^{D-n}\times T^n$, 
where compactification radii of $T^n$ are $L$ for all directions. 
Suppose that a Schwarzschild black hole is placed in this space. 
If its horizon is much smaller than $L$, then we can approximate it 
using a Schwarzschild solution in ${\mathbb R}^D$. 
If the size of the horizon becomes comparable to $L$, 
the black hole cannot fit in $T^n$ and can wind on some cycles   
$T^p$ in $T^n$. This is the Gregory-Laflamme transition. 
Preferred value of $p$ can be determined by comparing the free energy. 

Let us begin with the simplest case, $p=n$. 
In this case, the metric can be written as 
\begin{equation}\label{eq:bsTn}
ds^2 = 
-\left( 
1
- 
\left(
 \f{r_H^{(n)}}{r}\right)^{D-n-3}
\right)dt^2 
+ 
\f{dr^2}{1- \left( \f{r_H^{(n)}}{r}\right)^{D-n-3}}
+ r^2d\Om^2_{D-n-2} + \sum_{i=1}^n dy_i^2, 
\end{equation}
where $r_H^{(n)}$ represents the horizon radius 
and $0 \le y_i \le L ~(i=1\dots n)$. 
By requiring the regularity of the above metric, 
the Hawking temperature $T_H^{(n)}$ is given by 
\begin{equation}\label{eq:Hatem}
T_H^{(n)} = \f{D-n-3}{4\pi r_H^{(n)}}.
\end{equation}

The ADM energy (mass) and entropy of the black string (\ref{eq:bsTn}) 
are given by \cite{HaHo}
\begin{align}
M(n) &= \f{1}{16\pi G_N^{(D)}}(D-n-2)\Om_{D-n-2}(r_H^{(n)})^{D-n-3} L^n,  
\label{ADM energy_Schwarzschild}\\
S(n) &= \f{1}{4G_N^{(D)}}\Om_{D-n-2} (r_H^{(n)})^{D-n-2} L^n.
\label{entropy_Schwarzschild}
\end{align}

For $p<n$, because exact metric is not known, 
we approximate the transverse compact directions by noncompact ones. 
Then, the above expressions can be used by replacing $n$ with $p$. 
In several examples, 
this approximation is known to be good even if the horizon 
is comparable to the compactification radius; see e.g. \cite{KW}. 
In the present case, the ratio $r_H/L$ is around $0.3\sim 0.4$ and 
we can expect that this approximation is valid. 

\subsection{$p$-brane vs. $(p-1)$-brane}
In order to determine the phase structure, 
let us compare the free energies for $p$-branes and $(p-1)$-branes 
fixing temperature $T_H$ and varying $L$. 
Then, because the Hawking temperature is related to 
the size of the horizon as  
\begin{equation}\label{eq:temconst}
T_H^{(p)} = \f{D-p-3}{4\pi r_H^{(p)}} = T_H : {\rm independent\ of\ }p, 
\end{equation}
we have the relation 
\begin{equation}\label{eq:relhori}
r_H^{(p)} = \f{D-p-3}{D-p-2}r_H^{(p-1)}.
\end{equation}
The free energy of $p$-brane is given by
\begin{align}\label{eq:FreeEn}
F(p) &= M(p) - T_H^{(p)}S(p) 
= \f{\Om_{D-p-2}}{16\pi G_N^{D}} (r_H^{(p)})^{D-p-3}L^p.
\end{align}
When $F(p) > F(p-1)$, the $(p-1)$-brane is more favorable.
Using (\ref{eq:FreeEn}) and (\ref{eq:relhori}), this relation becomes 
\begin{equation}\label{eq:circond}
t\equiv T_H\cdot L 
>
 \left(\f{D-p-2}{D-p-3}\right)^{D-p-2}\f{\Om_{D-p-1}}{\Om_{D-p-2}}\f{D-p-3}{4\pi}
\equiv
 t_{C(p)}.
\end{equation}
As shown in Table \ref{tab:1}, the critical value $t_{C(p)}$ is the
decreasing function of $p$.

We can also compare the entropy with ADM energy fixed. 
Similar pattern of instability appears also in this case (see Table \ref{tab:1}). 

\begin{table}[htbp]
	\begin{center}
     \begin{tabular}{|c||c|c|}\hline
      &  mass fixed  & temperature fixed \\ \hline
      $ t_{C(1)} $ & 1.27 & 1.28 \\ \hline
      $ t_{C(2)} $ & 1.15 & 1.17 \\ \hline
      $ t_{C(3)} $ & 1.01 & 1.04\\ \hline
     \end{tabular}
     \caption{The critical values of the circle for ${\mathbb R}^{7}\times T^3$. }
     \label{tab:1}
    \end{center}
\end{table}

\begin{figure}[htbp]
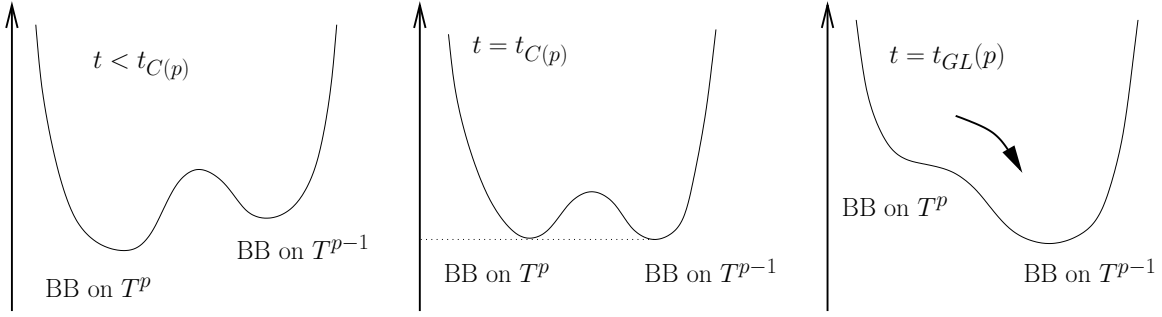

\begin{tabular}{ccc}
\begin{minipage}{5cm}
  \begin{center}
    \scalebox{0.4}{
      \input{Transition1.pstex_t}
    }
  \end{center}
\end{minipage}
&
\begin{minipage}{5cm}
  \begin{center}
    \scalebox{0.4}{
      \input{Transition2.pstex_t}
    }
  \end{center}
\end{minipage}
&
\begin{minipage}{5cm}
  \begin{center}
    \scalebox{0.4}{
      \input{Transition3.pstex_t}
    }
  \end{center}
\end{minipage}
\end{tabular}
\caption{A plausible picture of a phase transition. 
[Left] At $t>t_{C(p)}$, $p$-brane has smaller free energy and is favored. 
[Center]  At $t=t_{C(p)}$, free energies of $p$-  and $(p-1)$-branes become the same. 
[Right] At $t=t_{GL(p)}$, $p$-brane becomes unstable and decays to $(p-1)$-brane.}
\label{fig:1}
\end{figure}

The argument above does not exclude the possibility that several 
phases co-exist. That the instability really arises can be confirmed 
by calculating the Gregory-Laflamme (GL) critical temperature \cite{GrLa}, 
where $p$-branes become unstable thermodynamically \cite{GuMi,HuRa}. 
As we will show in the following, there is a relation\footnote{
The authors are grateful to J.~Marsano for instructive comments concerning 
the derivation of (\ref{RelationBetweenTcAndTGL}). 
} 
\begin{eqnarray}
t_{C(p)}
<
t_{GL(p)}
<
t_{C(p-1)}
\label{RelationBetweenTcAndTGL}
\end{eqnarray}
for $D=10$. (Note that this relation does not hold for large $D$.)
The relation (\ref{RelationBetweenTcAndTGL}) suggests that the $p$-brane 
decays to the $(p-1)$-brane at $t=t_{GL(p)}$ as depicted in Figure~\ref{fig:1}. 
Then it is plausible that the transition is of first order.

Now let us evaluate $t_{GL(p)}$. 
By approximating the compact directions transverse to $p$-brane 
by noncompact ones, the problem reduces to the evaluation of 
the GL temperature for $p$-brane wrapped on $T^p$ 
in ${\mathbb R}^{D-p}\times T^p$. This has been studied already in \cite{KS,Reall,HNO2} 
(see Table \ref{tab:1.5}). 
\begin{table}[htbp]
	\begin{center}
     \begin{tabular}{|c||c|}\hline
      $ t_{GL(1)} $ &  1.30\\ \hline
      $ t_{GL(2)} $ &  1.20\\ \hline
      $ t_{GL(3)} $ &  1.08\\ \hline
     \end{tabular}
     \caption{The Gregory-Laflamme critical temperature for ${\mathbb R}^{7}\times T^3$. }
     \label{tab:1.5}
    \end{center}
\end{table}

\begin{figure}[htbp]
\begin{tabular}{ccc}
\begin{minipage}{5cm}
  \begin{center}
      \includegraphics[scale=0.35,clip]{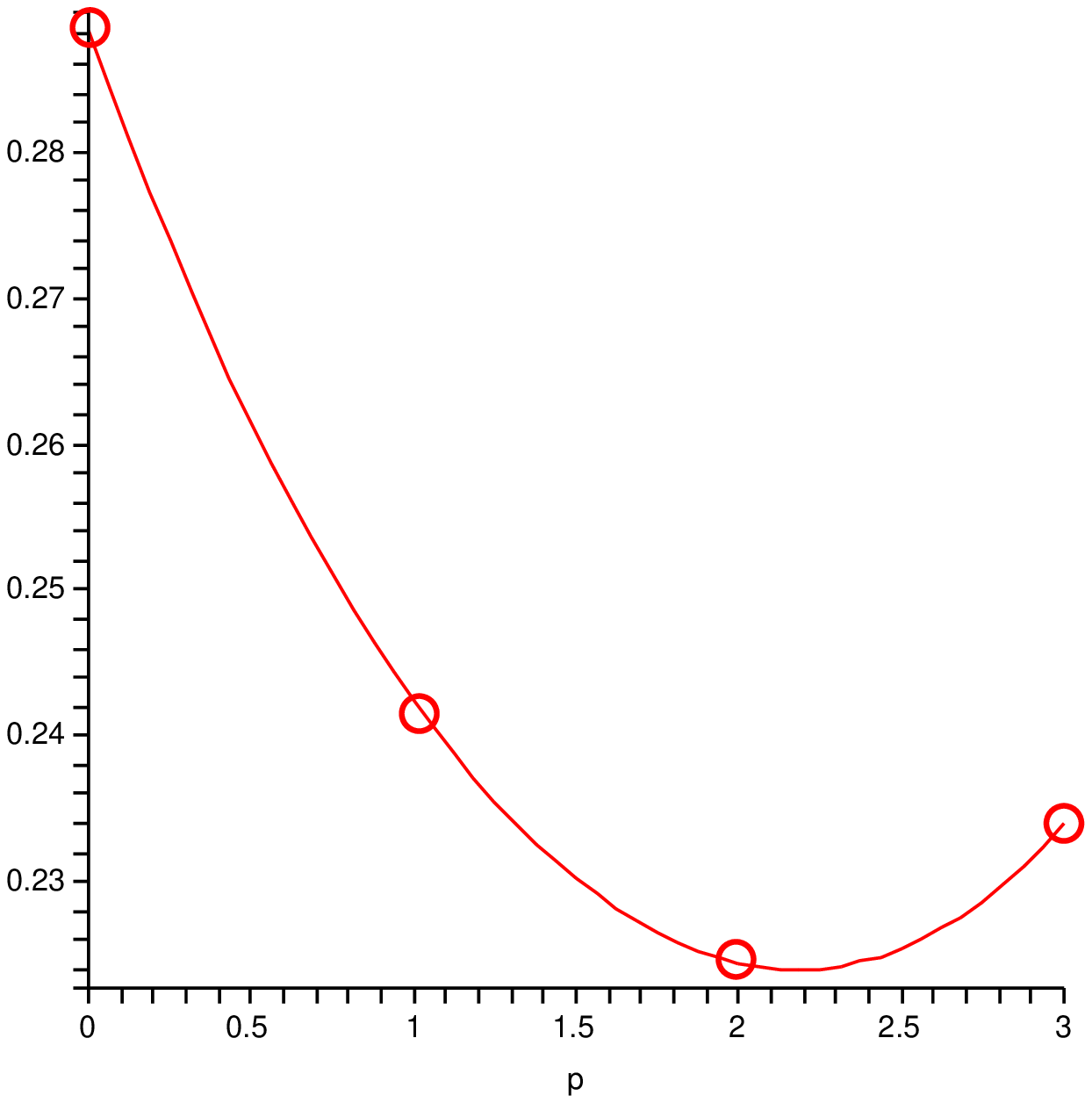}
  \end{center}
\end{minipage}
&
\begin{minipage}{5cm}
  \begin{center}
    \includegraphics[scale=0.35,clip]{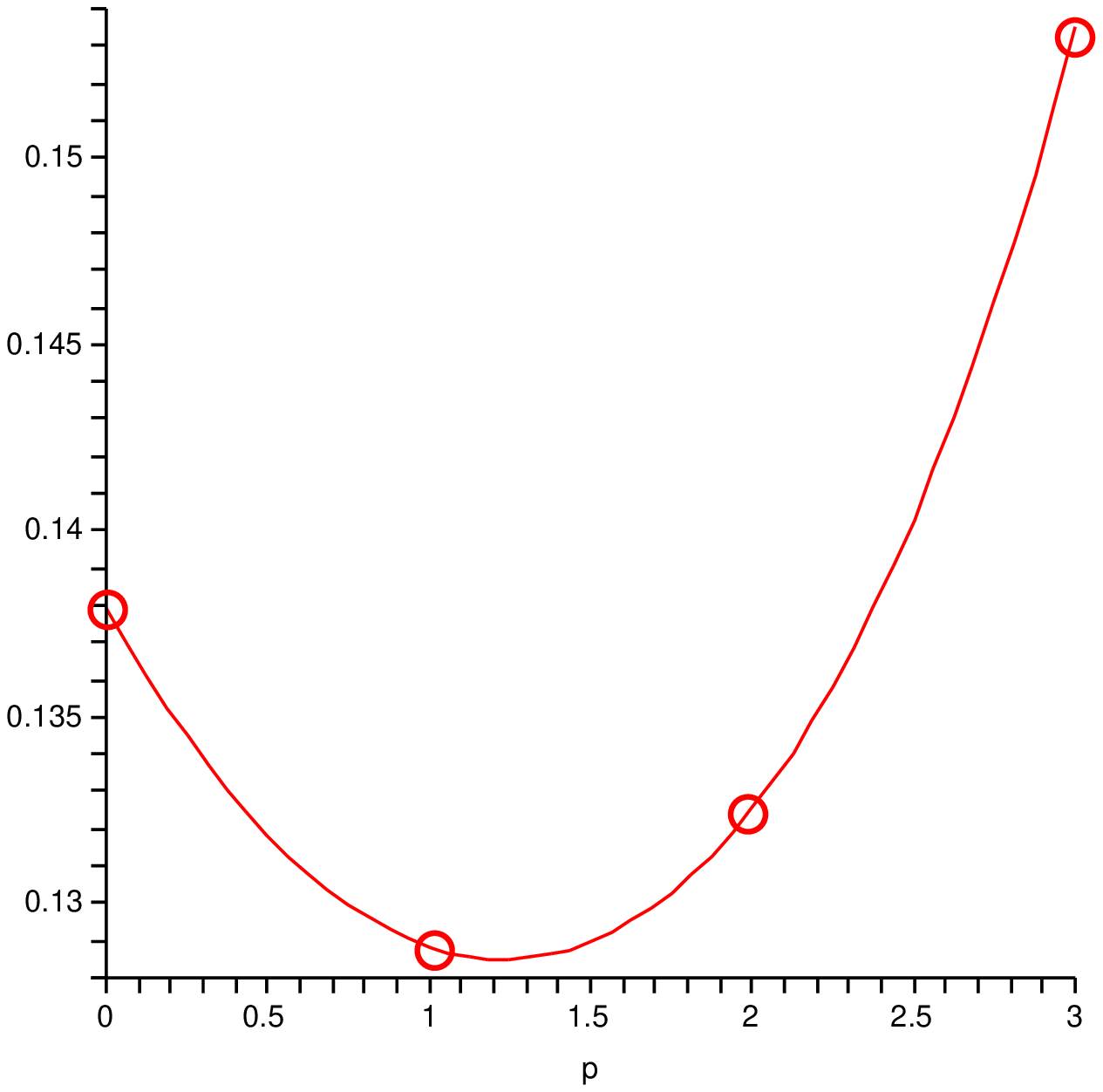}
  \end{center}
\end{minipage}
&
\begin{minipage}{5cm}
  \begin{center}
    \includegraphics[scale=0.35,clip]{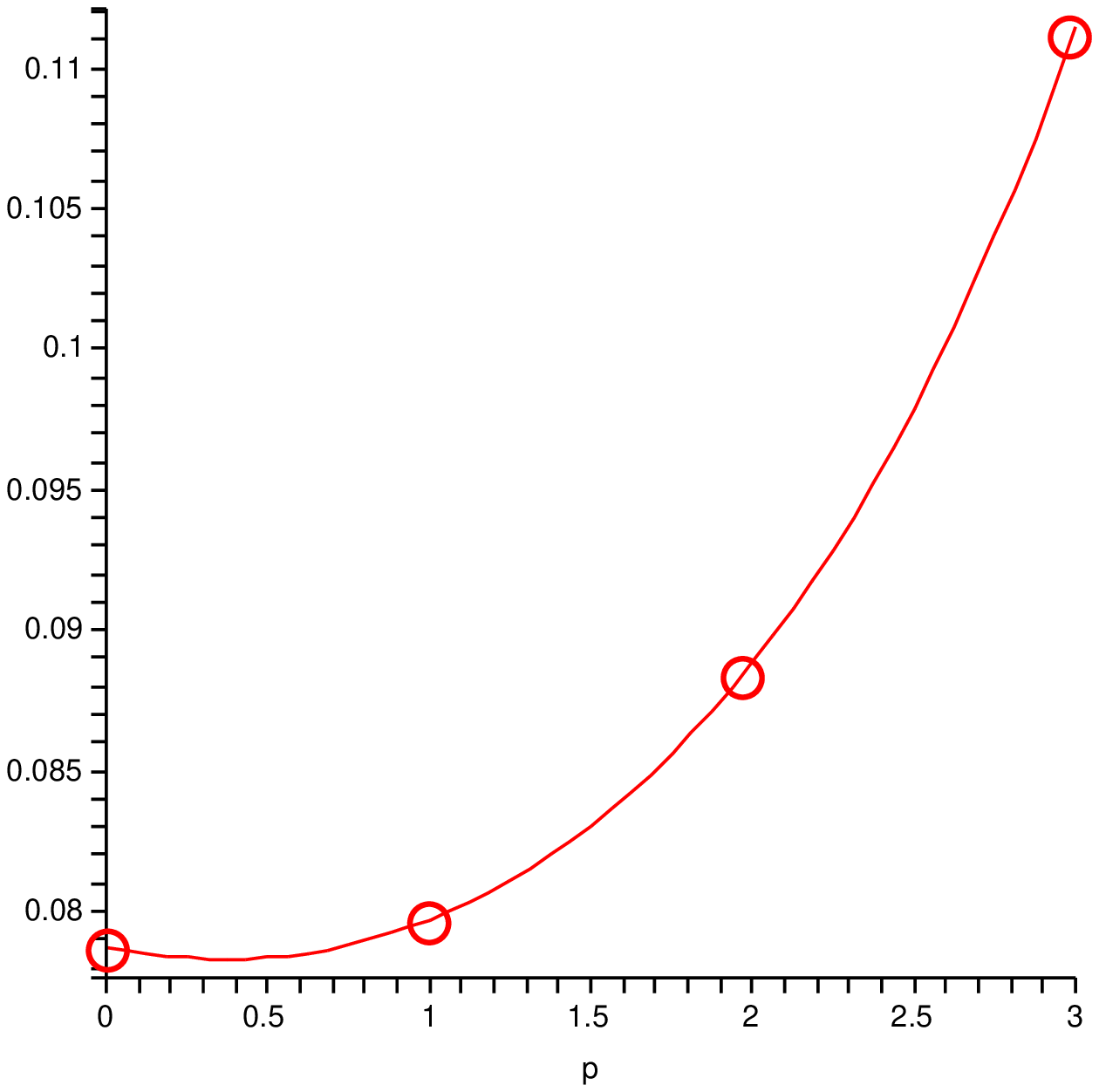}
  \end{center}
\end{minipage}
\end{tabular}
\caption{Comparison of free energies at $t_{C(p)}$ in unit of $L^{D-3}/16\pi G_N^{D}$ for $D=10, ~p\le 3$. 
Only $(p-1)$-brane has smaller free energy than $p$-brane and is favored.
[Left] $t=t_{C(3)}$.   
[Center]  $t=t_{C(2)}$. 
[Right] $t=t_{C(1)}$.
}
\label{fig:1.5}
\end{figure}

Furthermore, we expect $p$-brane decays not to $0$-brane directly, but to $(p-1)$-brane.
We check this below for $D=10, ~p\le 3$ by comparing free energies at $t_{GL(p)}$.
Figure \ref{fig:1.5} shows that only $(p-1)$-brane has smaller free energy than $p$-brane
at $t_{GL(p)}$. This result makes us confirm the above statement is true\footnote{
Similar cascade can be found when we compare the entropies of $p$-branes with fixed mass. 
For example, for $D=10, n=3$, we have 
$S(0)< S(3)< S(1)< S(2)$
at $t=t_{GL(3)}$. 
Therefore, 3-brane can decay only to 1- and 2-branes.
Entropically, 2-brane is most favorable.
Once 3-brane decays to 2-brane, then we have $S(2)<S(0)<S(1)$ at $t=t_{GL(2)}$,
and hence 2-brane can decay to 0- and 1-branes.
Again, 1-brane is more favorable entropically.
This result is consistent with the result in \cite{KS} .
}. 
Similar statement can be seen in \cite{KoSo2}, where $n$-brane on 
$T^n$ has been studied and it was shown that tachyonic mode 
develops along one of the $n$-directions \footnote{
We are grateful to E.~Sorkin for informing us \cite{KoSo2}.
}.     
It is important to study instability around $p$-brane $(p<n)$ in $T^n$ by metric 
perturbation and determine whether the tachyonic mode grows along one of the directions 
also in this case. Also it is necessary to calculate critical temperatures 
$t_{C(i)}$ and $t_{GL(i)}$ more 
precisely, because a few percent error can destroy the cascade\cite{KoSo2}.

Alternatively, we can compare the entropy while taking the mass of 
$p$- and $p-1$-branes to be equal. In this case, as $M$ decrease 
through radiation, $S(p)$ becomes smaller than $S(p-1)$ at 
\begin{eqnarray}
	M|_{p\to p-1}
	=
	\frac{L^{D-3}}{16\pi G_N^{D}}
	\frac{(D-p-2)^{(D-p-2)^2}}{(D-p-1)^{(D-p-1)(D-p-3)}}
	\frac{(\Omega_{D-p-2})^{D-p-2}}{(\Omega_{D-p-1})^{D-p-3}}. 
\end{eqnarray}
This $M|_{p\to p-1}$ is monotonically increasing with $p$,  
and hence transitions can take place repeatedly. 

\section{Super Yang-Mills theory on $T^{n+1}$ and its phase structure}
\label{sec:SYM}
\setcounter{equation}{0}
In this section, we consider the $U(N)$ super Yang-Mills theory 
on Euclidean torus $T^{n+1}\ (n=1,2,3)$.\footnote{
The argument below is a generalization of that in \cite{AMMW} to 
$n>1$. Here we take large-$N$ limit.
} The action is given by 
\begin{eqnarray}
S_{YM}
=
\frac{N}{4\lambda}
\int_0^{1/T_H}dt\int_0^L dx^1\cdots\int_0^L dx^n
\ Tr\left\{
F_{\mu\nu}^2
+
2(D_\mu\phi^I)^2
-
[\phi^I,\phi^J]^2
+
({\rm fermion})
\right\},  \nonumber\\
\end{eqnarray}
where $\mu=0,\cdots,n$, $x^0=t$, $\phi^I\ (I=1,\cdots,9-n)$ 
are adjoint scalars, and  fermions are anti-periodic along the thermal cycle. 

At high-temperature, KK modes along thermal cycle decouple and the model 
can be described by bosonic Yang-Mills on $T^n$. If $L$ is small, 
spatial KK modes decouple and the model reduces to BFSS matrix model \cite{BFSS}. 
When both thermal and spatial cycles are small, it reduces to 
a bosonic version \cite{HNT} of the IKKT matrix model \cite{IKKT}. 
In the following, we determine the parametric region where the above reduction 
is justified, and then discuss phase structure there.  

By rewriting the action using dimensionless quantities, we have 
\begin{eqnarray}
S_{YM}
=
\frac{N}{4\lambda^\prime}
\int_0^{1/t_H}dt\int_0^1 
dx^1\cdots\int_0^1 dx^n
\ Tr\left\{
F_{\mu\nu}^2
+
2(D_\mu\phi^I)^2
-
[\phi^I,\phi^J]^2
+
({\rm fermion})
\right\}, 
\nonumber\\ 
\end{eqnarray}
where 
\begin{eqnarray}
\lambda^\prime=\lambda L^{3-n}, 
\qquad
t_H=T_H L, 
\end{eqnarray}
and fields are also redefined to be dimensionless 
e.g. as $A^{new}_\mu= LA_\mu^{old}$.  
If spatial KK modes decouple before 
temporal KK modes do, then the action can be rewritten as 
\begin{eqnarray}
S_{YM}
=
\frac{N}{4\lambda^\prime}
\int_0^{1/t_H}dt
\ Tr\left\{
F_{\mu\nu}^2
+
\cdots
\right\}
=
\frac{N}{4}
\int_0^{\lambda^{\prime 1/3}/t_H}
d\tilde{t}
\ Tr\left\{
\tilde{F}_{\mu\nu}^2
+
\cdots
\right\}, 
\nonumber\\ 
\end{eqnarray}
where $\tilde{t}=\lambda^{\prime 1/3}t$ and 
$\tilde{F}=\lambda^{-\prime 2/3}F$.  
Therefore, temporal zero modes decouple if 
\begin{eqnarray}
t_H\gg\lambda^{\prime 1/3}.  
\end{eqnarray}   

In the same way, if temporal KK modes decouple before 
spatial KK modes do, then spatial KK modes decouple when   
\begin{eqnarray}
t_H\ll \frac{1}{\lambda^\prime}.  
\end{eqnarray}

In summary, SYM on $T^{1+n}$ can be described by bosonic IKKT model 
(i.e. can be reduced to zero-dimensional bosonic model) if 
\begin{eqnarray}
\lambda^{\prime 1/3}\ll t_H\ll \frac{1}{\lambda^{\prime}}. 
\label{validity of SYM}
\end{eqnarray}
In this region, both Polyakov loop and 
the spatial Wilson loops have nonzero expectation values. 
At $t_H\gtrsim \frac{1}{\lambda^{\prime}}$, the model reduces to 
bosonic YM on $T^p$ with $(10-p)$ adjoint scalars, 
which can be studied through lattice simulation. 
In this region, the Polyakov loop has nonzero expectation value 
and hence the model is in the black hole/black string phase.
Condensation of spatial Wilson loops takes place at 
$t_H\sim \frac{1}{\lambda^{\prime}}$. 

Let us consider the case of $n=2$.\footnote{$n=1$ case has been studied 
in \cite{AMMW}. Simulation results can be found in \cite{AMMW,JaWo,BiWo,KNT,KNT2}. } 
Then, at $t_H\gtrsim \frac{1}{\lambda^{\prime}}$, 
SYM is well approximated by two-dimensional YM with 8 adjoint scalars.  
Although a simulation of this model has not yet performed   
as far as we know, we can guess its properties from simulations 
on similar models. 
In \cite{NN}, bosonic pure $U(N)$ YM on $T^3$ and $T^4$ 
have been studied.  
By taking the sizes of one or two directions to be zero, they become 
bosonic YM on $T^2$ with one or two adjoint scalars, respectively.  
Let us take two compactification radii of $T^2$ to be the same value $L$. 
Then, the action is invariant under the exchange of two directions. 
Let the spatial Wilson loops to be 
\begin{eqnarray}
	W_\mu=\f{1}{N}\cdot Tr\left[ P\exp\left(i\int_0^L dx^\mu A_\mu\right) \right]\qquad (\mu=1,2),
\end{eqnarray}
where the indices $\m$ are not contracted in r.h.s.
Spatial Wilson loops have nonzero expectation values if 
the global $U(1)^2$ symmetry 
\begin{eqnarray}
A_\mu\to A_\mu+c_\mu\cdot\textbf{1}_N
\end{eqnarray}
is spontaneously broken.  
As discussed in \cite{NN}, the $U(1)^2$ symmetry is actually broken and 
$\langle W_\mu\rangle$ behaves as 
\begin{eqnarray}
& &
\langle W_1\rangle
=
\langle W_2\rangle
=
0
\qquad
(L\ge{}^\exists L_c^{(1)}), 
\\
& &
\langle W_1\rangle
\neq 0,\quad
\langle W_2\rangle
=
0
\quad
{\rm or}
\quad
\langle W_2\rangle
\neq 0,\quad
\langle W_1\rangle
=
0
\qquad
({}^\exists L_c^{(2)}\le L\le{}^\exists L_c^{(1)}), 
\\
& &
\langle W_1\rangle
\neq
0,
\langle W_2\rangle
\neq
0
\qquad
(L\le{}^\exists L_c^{(2)}). 
\end{eqnarray}
We can expect that this pattern does not 
depend on the number of adjoint scalars. 
We can also expect that the result is similar also for other values of $n$. 
Note that these transitions are likely to be of first order. 

Suppose that such a pattern of phase transition persists to 
strong coupling region.  
In the dual gravity theory, the eigenvalue 
distribution of spatial Wilson loops is related 
to that of D$0$-branes in T-dual picture, and 
condensation of spatial Wilson loop corresponds to 
the Gregory-Laflamme transition in D$0$-brane system 
\cite{Su,BKR,MaSa,MaSa2,AMMW,HO}. 
Therefore, transitions in super Yang-Mills theory can be regarded 
as a sequence of Gregory-Laflamme transitions in gravity side. 
In the next section, we show such successive transitions do exist 
in supergravity. 
\section{Cascade of Gregory-Laflamme transitions in IIA Supergravity}
\label{sec:gravity}
\setcounter{equation}{0}
In this section,
we study the dual gravity theory of D$p$-branes on $T^n$ corresponding to 
the super Yang-Mills theory on square torus discussed in the previous section
 and find an indication for the cascade.  
As we will show, the transitions take place in the parametric region where the 
T-dual picture, a system of D$0$-branes, is valid.

\subsection{Generalities}
We consider the
near-extremal D$p$-brane solution on $T^n$ which is dual
to the super Yang-Mills theory on $T^n$ in finite temperature. 
As in \S\ref{sec:Scwarzschild}, for $p<n$  
we approximate transverse compact dimensions by noncompact ones. 

In the near horizon limit the metric and dilaton are given by \cite{IMSY,HoSt}
\begin{align}\label{eq:bbTp}
ds^2 = \al\Bigg\{ \f{U^{\f{7-p}{2}}}{g_{YM}\s{d_pN}}
&\left[
  -\left( 1- \f{U_0^{7-p}}{U^{7-p}}\right)dt^2 + \sum_{i=1}^pdy_i^2 \right] \\
 +& \f{g_{YM}\s{d_pN}}{U^{\f{7-p}{2}}\left( 1- \f{U_0^{7-p}}{U^{7-p}}\right)}dU^2 
+ g_{YM}\s{d_pN}U^{\f{p-3}{2}}d\Om_{8-p}^2 \Bigg\},\nonumber\\
e^\phi = (2\pi)^{2-p}g^2_{YM}&\left( \f{g_{YM}^2 d_pN}{U^{7-p}}
\right)^{\f{3-p}{4}},\qquad d_p = 2^{7-2p}\pi^{\f{9-3p}{2}}\G\left(\f{7-p}{2}\right), 
\end{align}
where $0\le y_i\le L~(i=1,\cdots, p)$ denotes the direction of $T^p$. 
The Yang-Mills coupling constant is related to the string coupling at the spatial 
infinity $g_s=e^{\phi_\infty}$ as 
\begin{equation}\label{eq:YMandString}
g_{YM}^2 = (2\pi)^{p-2}g_s\alpha^{\prime \f{p-3}{2}}.
\end{equation}
Requiring the regularity of the metric around
the horizon $r_H$, the Hawking temperature is determined to be  
\begin{equation}\label{eq:HawkingTempOfDp}
  T_H = \f{(7-p)U_0^{\f{5-p}{2}}}{4\pi\s{d_p\l}}, 
\end{equation}
where $\l\equiv g_{YM}^2N$. 
Using the Einstein frame the ADM energy and the entropy of this
solution can be computed as \cite{IMSY} 
\begin{align}\label{eq:ADMmass}
E_{Dp} &= \f{9-p}{2^{11-2p}\pi^{\f{13-3p}{2}}
\G (\f{9-p}{2})\l^2}N^2U_0^{7-p}L^p,\\
S_{Dp} &= \f{1}{2^{8-2p}\pi^{\f{11-3p}{2}} \G (\f{9-p}{2}) \l^2}\s{d_p\l}N^2U_0^{\f{9-p}{2}}L^p.\label{eq:DpEntropy}
\end{align}

We can use the metric (\ref{eq:bbTp}) as a supergravity solution 
when the string excitation and winding modes wrapping on $T^p$ 
are much heavier than KK modes along $S^{8-p}$:
\begin{align}
&\f{1}{\s\l L} \ll T_H , \qquad 1 \ll \l \qquad (p=3)\\ 
&\f{1}{\s\l L^{(5-p)/2}} \ll T_H \ll \l^{\f{1}{3-p}} \qquad (p=0,1,2).
\end{align}
\begin{flushleft}
{\large \it T-dual picture}
\end{flushleft}
By taking a T-dual in all the compact directions of $T^p$,
the metric and dilaton become
\begin{align}\label{eq:bbTpT}
ds^2 =& \al\Bigg\{ -\f{U^{\f{7-p}{2}}}{\s{d_p\l}}\left( 1-
  \f{U_0^{7-p}}{U^{7-p}}\right)dt^2
+ \s{d_p\l}U^{\f{p-3}{2}}d\Om_{8-p}^2  \nonumber\\
& \qquad \qquad \qquad  + \f{\s{d_p\l}}{U^{\f{7-p}{2}}}\left[
\f{dU^2}{\left( 1- \f{U_0^{7-p}}{U^{7-p}}\right)} + \sum_{i=1}^pd\ti y_i^2
\right] \Bigg\},\\
 e^\phi =& (2\pi)^2\f{\l}{N}\left( \f{d_p\l}{U^{7-p}} \right)^{\f{3}{4}}\f{1}{L^p},
\end{align}
where $\ti y_i$ denotes the T-dualized circle coordinate and takes the
range\footnote{
We have used the relation between 
the dilaton before and after the T-dual 
$e^{\ti \phi} = e^\phi \f{\alpha^{\prime 1/2}}{R}$, 
where $2\pi R = L$. 
} $0 \le \ti y_i \le (2\pi)^2/L$. 
The metric after T-dual represents the uniform distribution of $N$ D0-branes on $T^p$.
The Hawking temperature, ADM energy and entropy are the same 
as those before taking T-dual, 
(\ref{eq:HawkingTempOfDp}), (\ref{eq:ADMmass}) and (\ref{eq:DpEntropy}). 
Winding modes are negligible when 
\begin{equation}\label{condition for sugra approximation1}
T_H L\ll 1,
\end{equation}
and the condition that the $\al$ correction can be neglected 
does not change:  
\begin{equation}\label{condition for sugra approximation2}
T_H \ll \l^{1/(3-p)} \quad (p=0,1,2) , \qquad 1 \ll \l \quad (p=3). 
\end{equation}
\subsection{D$p$ vs. D$(p-1)$}
From (\ref{eq:ADMmass}) and (\ref{eq:DpEntropy}), the free energy
of $N$ D$p$-branes is given by 
\begin{equation}
F(p)= -\f{5-p}{2^{11-2p}\pi^{\f{13-3p}{2}}
\G (\f{9-p}{2})\l^2}N^2U_0^{7-p}L^p.
\end{equation}
By using dimensionless parameters 
$t \equiv T_HL,~u_0\equiv U_0L$ and $\l'\equiv \l L^{3-p}$,  
ADM energy, entropy and free energy can be rewritten as   
\begin{align}\label{eq:FreeEnergyDp}
E(p) &= \f{(9-p)B(p)}{L} N^2\l_{(p)}^{\prime \f{p-3}{5-p}}t_{(p)}^{\f{2(7-p)}{5-p}}, \\
S(p) &= 8\pi B(p) N^2\s{d_p\l'}\l_{(p)}^{\prime \f{p-3}{5-p}}t_{(p)}^{\f{9-p}{5-p}},\\
F(p) &= -\f{(5-p)B(p)}{L} N^2\l_{(p)}^{\prime \f{p-3}{5-p}}t_{(p)}^{\f{2(7-p)}{5-p}},\label{eq:FreeEnDp}\\ 
t_{(p)} = \f{(7-p)u_0^{\f{5-p}{2}}}{4\pi\s{d_p\l'}}&, \qquad 
B(p) \equiv \f{1}{2^{11-2p}\pi^{\f{13-3p}{2}}\G (\f{9-p}{2})}\left( \f{4\pi\s{d_p}}{7-p}\right)^\f{2(7-p)}{5-p}.\nonumber
\end{align}
Here $(p)$ indicates that these quantities depend on $p$ in general.  
Note, however, that the effective 't Hooft coupling $\l'_{(p)}$ 
is indeed independent of $p$. 
This can be seen as follows. 
We can write the 't Hooft coupling on $T^p$ using (\ref{eq:YMandString})
\footnote{We 
return to the metric before taking near-horizon limit.} 
\begin{equation}
\l_{(p)} = g_{YM}^2N = (2\pi)^{p-2}e^{\phi_\infty^{(p)}}\alpha^{\prime \f{p-3}{2}}.
\end{equation}
Since the string coupling on the $T^p$ and its T-dual is related as 
\begin{equation}
e^{\phi^{(0)}_\infty} = e^{\phi^{(p)}_\infty}\left( \f{\s{\alpha^\prime}}{L/2\pi} \right)^p, 
\end{equation}
we have 
\begin{equation}
\l_{(p)} = \l_{(0)}L^p.
\end{equation}
Then it is clear the effective coupling does not depend on $p$: 
\begin{equation}
\l'_{(p)}=\l' = \l_{(0)}L^3.
\end{equation} 
\begin{flushleft}
{\large \it Temperature fixed comparison}
\end{flushleft}
We compare $F(p)$ and $F(p-1)$ at the same $T_H$ and $L$ 
(and hence $t$) in the strong coupling region $\lambda^\prime\gg 1$.

Using (\ref{eq:FreeEnDp}) we can easily see $F(p) > F(p-1)$ 
is realized when the dimensionless temperature $t$ becomes lower than the critical
temperature $t_{C(p)}$, 
\begin{align}\label{eq:TempFixTr}
t &< t_{C(p)} \equiv \f{A(p)}{\s{\l'}}, \qquad
A(p) = \left( \f{(6-p)B(p-1)}{(5-p)B(p)} \right)^{\f{(5-p)(6-p)}{4}}.
\end{align}
As shown in Table \ref{tab:2}, $t_{C(p)}$ is the increasing function of $p$.
\begin{table}[htbp]
	\begin{center}
	 \begin{tabular}{|c||c|}\hline
      $t_{C(4)}$ & $2.92 /\sqrt{\lambda^\prime}$ \\ \hline
      $t_{C(3)}$ & $2.87 /\sqrt{\lambda^\prime}$ \\ \hline
      $t_{C(2)}$ & $2.67 /\sqrt{\lambda^\prime}$ \\ \hline
      $t_{C(1)}$ & $2.40 /\sqrt{\lambda^\prime}$ \\ \hline
     \end{tabular}
     \caption{The critical effective temperatures}
     \label{tab:2}
    \end{center}
\end{table}

These critical points lie in the parametric region 
(\ref{condition for sugra approximation1}) 
where D$0$-brane picture is valid. 
In this region 
KK modes along $T^n$ can also be neglected. 
Note that, in this picture, smaller $t$ corresponds to larger compactification radius $\ti L \sim 1/L$.
Therefore, the relation 
\begin{equation}
t_{C(1)} < \dots < t_{C(p-1)} < t_{C(p)} 
\end{equation}
indicates that for large $\ti L$ low dimensional object is favored similarly to the Schwarzschild case.

\begin{figure}[htbp]
\begin{center}
\includegraphics[width=8cm,clip]{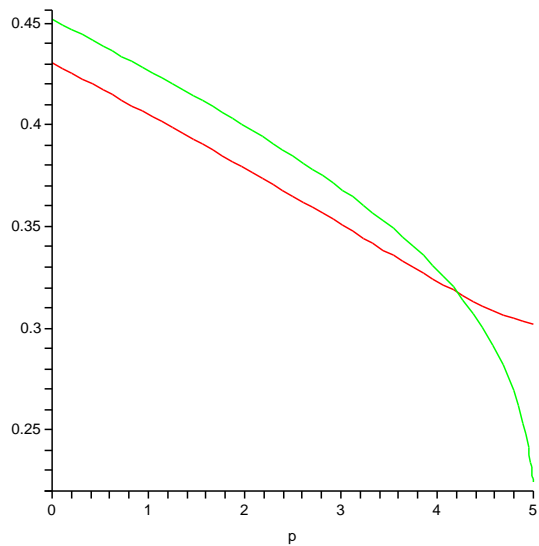}
\caption{The plot of $a(p-1)$ (red) and $b(p)$ (green).}
\label{fig:compare}
\end{center}
\end{figure}

Above we have assumed that the non-uniform phase where D$0$-branes 
collapse in some circle directions
can be approximated by those on flat noncompact space.  
In several examples, 
this assumption is known to be valid when the horizon is smaller than 
the radius of the circle.
In the present case, the horizon of the D$0$-branes on $T^p$ and the radius of the T-dualized torus 
before taking near-horizon limit are given as $r_H^{(p)} \equiv u_0^{(p)}\al /L$ and
$\ti L \equiv (2\pi)^2\al /L$. 
Let us consider the ratio at the critical temperature $t_{C(p)}$
\begin{equation}
a(p) \equiv \f{r_H^{(p)}}{\ti L}\Big|_{t = t_{C(p)}} = \f{1}{(2\pi)^2}\left( \f{4\pi\s{d_p}A(p)}{7-p} \right)^\f{2}{5-p}.
\end{equation}
After the transition D$p$-branes become D$(p-1)$-branes and the radius of the horizon changes to 
\begin{equation}
b(p) \equiv \f{r_H^{(p-1)}}{\ti L}\Big|_{t = t_{C(p)}} = \f{1}{(2\pi)^2}\left( \f{4\pi\s{d_{p-1}}A(p)}{8-p} \right)^\f{2}{6-p}.
\end{equation}
Above ratios take values around $0.3 \sim 0.4$ 
and we can expect that the assumption is valid (see Table~\ref{tab:3} and 
Figure~\ref{fig:compare}).
\begin{table}[htbp]
	\begin{center}
	 \begin{tabular}{|c||c|c|}\hline
      $p$ & $a(p)$ & $b(p)$\\ \hline
       4 & 0.302 & 0.329  \\ \hline
       3 & 0.323 & 0.369  \\ \hline
       2 & 0.351 & 0.400  \\ \hline
       1 & 0.379 & 0.427  \\ \hline
     \end{tabular}
     \caption{The ratio between the horizon and the torus radius at the critical temperature.}
     \label{tab:3}
    \end{center}
\end{table}
We can also see that $a(p-1) < b(p)$ for $p\le 4$. 
This relation indicates that, 
after a transition from $p$-brane to $(p-1)$-brane, 
the horizon of $(p-1)$-brane is larger than the next critical radius  
$r_H^{(p-1)}|_{t=t_{C(p-1)}}$ 
and hence the next transition can take place when the horizon 
shrinks to $r_H^{(p-1)}|_{t=t_{C(p-1)}}$. 
In this way, transitions take place repeatedly.  

In the present case, we have not calculated the GL temperature $t_{GL(p)}$. 
In \cite{AMMW}, $t_{GL(1)}$ has been calculated, and the result is 
\begin{eqnarray}
t_{GL(1)}=\frac{2.29}{\sqrt{\lambda^\prime}}. 
\end{eqnarray} 
This is slightly lower than $t_{C(1)}$ as expected. 
If other $t_{GL(p)}$'s satisfy 
\begin{eqnarray}
t_{C(p-1)}
<
t_{GL(p)}
<
t_{C(p)}, 
\label{ConditionForCascadeInAdS}
\end{eqnarray} 
then a cascade can take place as first order transitions 
similarly to the case of dual SYM. 
Such a cascade resembles also to the Schwarzschild case.

\section{Discussion}

There are several directions for future studies. 
First of all, it is important to justify the approximation in this paper
 in which compact directions are replaced with noncompact ones, 
because phase structure is sensitive to the small changes of physical
quantities; as discussed in \cite{KoSo2}, a few percent error in the entropy 
can change the magnitude relation of critical temperatures and lead us to the wrong 
conclusion. At least in the case of D-brane discussed in \S 4, AdS/CFT correspondence 
suggests that this approximation gives the correct phase diagram.
 Secondly similar calculations for other compactifications 
are important. If compatified dimensions are curved, 
our approximation would not be valid at all, 
and more careful treatment would be necessary.  
In addition, it is necessary to consider other solutions 
such as a non-uniform string solution; at present, we showed 
that uniform $p$-brane solution is unstable and $(p-1)$-brane is 
(meta-)stable at a critical temperature, but it might be possible that $p$-brane decays to 
another stable solution. 
Also, it will be nice if it can be determined whether 
(\ref{ConditionForCascadeInAdS}) holds or not. 


It is possible to perform a similar study in gravity theories 
with other boundary conditions. 
From the point of view of the super Yang-Mills theory, 
such a study will be useful to clarify the phase structure of SYM 
on $T^{p+1}$ \cite{AMMPRW}.\footnote{The authors would like to thank J.~Marsano 
for pointing out this issue.} 

In large-$N$ gauge theories, if the global $U(1)$ symmetry 
is not broken then expectation values of Wilson loops are independent 
of compactification radii through Eguchi-Kawai equivalence \cite{EK}.  
Using this property, simulation cost could be reduced to a large extent \cite{NN}. 
Determination of the $U(1)$-unbroken region is important also for this reason.

\vskip .5in
\centerline{\bf Acknowledgments} 
The authors are grateful to T.~Azeyanagi, K.~Furuta, Y.~Matsuo,
 K.~Murata, K.~Oda, T.~Tanaka, T.~Takayanagi and especially 
to J.~Marsano and N.~Tanahashi for helpful discussions and comments. 
M.~H. is supported by Special Postdoctoral Researchers 
Program at RIKEN. 
 T.~N. would like to 
thank the Japan Society for the Promotion of Science for financial
support.

\vskip .3in
\newpage 


\end{document}